# Stability analysis of delayed system using Bode Integral

Anish Acharya[1], Debatri Mitra[2]

1. Department of Instrumentation and Electronics Engineering, Jadavpur University, Salt-Lake Campus, LB-8, Sector 3, Kolkata-700098, India.
2. Department of Electronics and communication Engineering, Future Institute of Engineering and Management
Email: anishacharya91@gmail.com

Kaushik Halder[3]

3. Department of Electronics and Instrumentation Engineering, National Institute of Science and Technology, Berhampur, India

*Abstract*— **The PID controller parameters can be adjusted in such a manner that it gives the desired frequency response and the results are found using the Bode's integral formula in order to adjust the slope of the nyquist curve in a desired manner. The same idea is applied for plants with time delay. The same has also been done in a new approach. The delay term is approximated as a transfer function using Padé approximation and then the Bode integral is used to determine the controller parameters. Both the methodologies are demonstrated with MATLAB simulation of representative plants and accompanying PID controllers. A proper comparison of the two methodologies is also done. The PID controller parameters are also tuned using a real coded Genetic Algorithm (GA) and a proper comparison is done between the three methods.**

*Keywords- PID controller, Ziegler-Nichols method, auto-tuning, Padé approximation, Bode's integral*

## I. INTRODUCTION

More than 90% of the controllers used in process industry are PID controllers[1]. However the performance of the PID controller like satisfactory tracking, disturbance rejection, robustness etc depends largely on the tuning methodology. Zieglar-Nichols methods[2] where the controller parameters are tuned from the knowledge of critical frequency and gain of the plant. In 1984 an automatic tuning method was proposed in [3] using simple relay feedback test which gives using the describing function analysis. However in order to obtain the desired phase and gain margin the closed loop system should oscillate at the desired crossover frequency which can be obtained using relay with hysteresis[4]or time delay which should slowly change up to obtain a limit cycle at the crossover frequency. However the experiment is more time consuming as compared to the previous one. A closed loop relay test was proposed in[5]where the crossover frequency is directly identified. In this method the plant operates in a closed loop with the controller and the output is fed back to the reference of the closed loop system. Karimi et. al.[6,7,8] used the Bode's integral[9] to approximate the derivatives of amplitude and phase of the system with respect to frequency at a given frequency. Then these derivatives have been used in the modified Ziegler-Nichols method in order to adjust the slope of the Nyquist curve at the given frequency.

The main objective of this paper to find the controller parameters in order to obtain desired frequency response for plants with pure time delay. The controller parameters have been tuned using the approximated derivatives of amplitude and phase with respect to frequency at the given frequency obtained using Bode's integral[9] as proposed by Karimi et. al.[6,7,8]. In this paper a new method is being proposed where the delay term is approximated using Padé approximation which replaces the delay term by a pure transfer function and then the controller parameters are tuned in order to obtain the desired frequency response using the approximated derivatives of amplitude and phase found directly from the Bode's integral formula. The responses of both the methods are compared with ample simulation examples.

The PID controller parameters are also found using a real coded Genetic Algorithm (GA). It is quite understandable that though individual variation in the controller parameters is smooth the combined landscape may be quite complicated consisting of local minima. This motivates the application of population based intelligent algorithm like GA over the classical gradient based methods in order to obtain the true global minima in the optimization process. The upper bound and lower bound are chosen in such a manner that it ensures much better closed loop performance with less error in the resultant slope of the Nyquist curve. Basically a trade off design is being proposed here between the closed loop performance and the resultant Nyquist slope. Thus this paper aims in proposing a new GA based PID controller tuning method for time delay systems resulting in a much better closed loop response as compared to the one proposed by Karimi et. Al. [6-8]

The rest of the paper is organized as follows. Section II describes adjustment of loop slope using Bode's gain phase relationship. Section III describes parameter tuning of the PID controller in order to obtain the desired response. In Section IV the effect of time delay on the parameters of the Controller found by directly applying the Bode's integral as described in section III. Section V presents a brief overview of Padé approximation which approximated the delay term . Ample simulation examples using MATLAB [10] are given in section VI for both the methods . Paper ends with the conclusion and scope of future work in section VII followed by references.

## II. LOOP SLOPE ADJUSTMENT USING BODE'S GAIN-PHASE RELATION

The robustness and performance of a closed loop system depends immensely on the slope of the nyquist curve at the crossover frequency. A relation between $T_i$ and $T_d$ of the controller can be derived in order to obtain the desired slope of the Nyquist curve of the loop transfer function at a given frequency. Consider the loop transfer function as

$$L(j\omega) = G(j\omega)K(j\omega) \quad (1)$$

Where the PID controller is given by

$$K(j\omega) = K_P(1 + \frac{1}{j\omega T_i} + j\omega T_d) \quad (2)$$

The slope of the nyquist curve of $L(j\omega)$ at $\omega_0$ is defined as $\psi$ which is equal to the phase of the derivative of $L(j\omega)$ at $\omega_0$. Then it can be derived that,

$$\psi = \varphi_0 + \tan^{-1}(\frac{(T_d T_i \omega_0^2 + 1) + (T_d T_i \omega_0^2 - 1)s_a(\omega_0) + s_p(\omega_0)T_i\omega_0}{s_a(\omega_0)T_i\omega_0 - (T_d T_i \omega_0^2 - 1)s_p(\omega_0)}) \quad (3)$$

Where,

$$\varphi_0 = \angle G(j\omega_0) \quad (4)$$

Further, $s_a(\omega_0)$ and $s_p(\omega_0)$ are defined as

$$s_a(\omega_0) = \omega_0 \frac{d\ln|G(j\omega)|}{d\omega}\bigg|_{\omega_0} \quad (5)$$

$$s_p(\omega_0) = \omega_0 \frac{d\angle G(j\omega)}{d\omega}\bigg|_{\omega_0} \quad (6)$$

$$T_d = \frac{s_a(\omega_0) - 1 + s_p(\omega_0)\tan(\psi - \varphi_0) - T_i\omega_0(s_p(\omega_0) - s_a(\omega_0)\tan(\psi - \varphi_0))}{\omega_0^2 T_i(1 + s_a(\omega_0) + s_p(\omega_0)\tan(\psi - \varphi_0))} \quad (7)$$

Now, $s_a(\omega_0)$ and $s_p(\omega_0)$ are approximated directly using the phase gain relationship reported by Bode[1] in 1945. The results are based on Cauchy's Residue theorem and have been used extensively in network analysis. There are two integrals presented [1]. The first integral shows the relation between the phase of the system at each frequency as a function of the derivative of its magnitude part and the second integral shows the relationship between amplitude of the system at any frequency to the derivative of phase and the static gain of the system.

The first relationship is given as:

$$\angle G(j\omega_0) = \frac{1}{\pi}\int_{-\infty}^{+\infty}\frac{d\ln|G(j\omega)|}{dv}\ln\coth\frac{|v|}{2}dv \quad (8)$$

Where, $v = \ln\frac{\omega}{\omega_0}$ since $\ln\coth\frac{|v|}{2}$ decreases rapidly as $\omega$ deviates from $\omega_0$ hence the integral depends mostly on $\frac{d\ln|G(j\omega)|}{dv}$ Hence it can be seen that,

$$\angle G(j\omega_0) \approx \frac{\pi}{2}\int_{-\infty}^{+\infty}\frac{d\ln|G(j\omega)|}{dv}dv\bigg|_{\omega_0} \quad (9)$$

Which is directly followed by the conclusion that

$$s_a(\omega_0) \approx \frac{2}{\pi}\angle G(j\omega_0) \quad (10)$$

Another relationship between gain and phase of a stable non minimum phase system is given by,

$$\ln|G(j\omega_0)| = \ln|K_g| - \frac{\omega_0}{\pi}\int_{-\infty}^{+\infty}\frac{d(\angle G(j\omega)/\omega)}{dv}\ln\coth\frac{|v|}{2}dv \quad (11)$$

Using (10) it can be concluded that,

$$s_p(\omega_0) \approx \angle G(j\omega_0) + \frac{2}{\pi}\left[\ln|K_g| - \ln|G(j\omega_0)|\right] \quad (12)$$

## III. PID CONTROLLER TUNING FOR DESIRED RESPONSE

In order to obtain a desired phase margin $\Phi_d$ at the crossover frequency the controller parameters can be found by adjusting the phase margin and the slope of the nyquist curve at desired the cross over frequency.

$$\angle G(j\omega_c) + \angle K(j\omega_c) = \Phi_d - \pi \quad (13)$$

$$|G(j\omega_c)K(j\omega_c)| = 1 \quad (14)$$

The parameters of the PID controller required in order to adjust the slope of the Nyquist curve in the desired way for improved system performance directly follows by solving (13), (14) and (7). The PID controller parameters thus found to be:

$$K_p = \frac{|(\cos(\Phi_d - \varphi_c))|}{|G(j\omega_c)|} \quad (15)$$

$$T_d = \frac{1}{2\omega_c}[(s_a(\omega_c) - s_p(\omega_c)\tan(\Phi_d - \varphi_c))\tan(\psi - \varphi_c)$$
$$+ (1 - s_a(\omega_c))\tan(\Phi_d - \varphi_c) - s_p(\omega_c)] \quad (16)$$

$$T_i = \frac{1}{\omega_c(T_d\omega_c - \tan(\Phi_d - \varphi_c))} \quad (17)$$

## IV. EFFECT OF TIME DELAY:

Suppose the system with pure time delay is described as
$$\mathbf{G}_\tau(j\omega) = G(j\omega)e^{-j\tau\omega} \quad (18)$$
Where $G(j\omega)$ is a stable non minimum phase system. Thus differentiating the amplitude and phase with respect to the frequency it can be shown that,
$$\frac{d\ln|G_\tau(j\omega)|}{d\omega} = \frac{d\ln|G(j\omega)|}{d\omega} \quad (19)$$
$$\frac{d\angle G_\tau(j\omega)}{d\omega} = \frac{d\angle G(j\omega)}{d\omega} - \tau \quad (20)$$
Thus it can be obtained using the Bode's gain phase relationship (9) that,
$$\angle G_\tau(j\omega_0) \approx \frac{\pi}{2}\left.\frac{d\ln|G(j\omega)|}{d\omega}\right|_{\omega_0} - \tau\omega_0 \quad (21)$$
Thus $s_a(\omega_0)$ and $s_p(\omega_0)$ can be found to be,
$$s_a(\omega_0) \approx \frac{2}{\pi}(\angle G_\tau(j\omega_0) + \tau\omega_0) \quad (22)$$
$$s_p(\omega_0) \approx \angle G_\tau(j\omega_0) + \frac{2}{\pi}\left[\ln|K_g| - \ln|G_\tau(j\omega_0)|\right] \quad (23)$$

## V. PADE APPROXIMATION FOR TIME DELAY SYSTEM

Padé approximation is frequently used to approximate a pure time delay by a rational transfer function. Thus on reducing the time delay term to a rational transfer function PID controller parameters may be adjusted using (15),(16),(17) for obtaining the desired response. The Padé approximation for the term $e^{-sL}$ is given by,
$$e^{-sL} \cong \frac{N_r(sL)}{D_r(sL)} \quad (24)$$
Where $N_r(sL)$ and $D_r(sL)$ are respectively given as,
$$N_r(sL) = \sum_{k=0}^{r}\frac{(2r-k)!}{k!(r-k)!}(-sL)^k \quad (25)$$
$$D_r(sL) = \sum_{k=0}^{r}\frac{(2r-k)!}{k!(r-k)!}(sL)^k \quad (26)$$

## VI. SIMULATION AND RESULTS

As a representative case the plant is chosen to be
$$G(s) = \frac{1}{(s+1)^5}e^{-0.1s} \quad (27)$$
Here the specifications are set as $0.4$ rad/sec as the desired crossover frequency and $50^0$ as the desired phase margin. Thus it is required to find the parameters of the PID controller which is capable of improving the closed loop performance of the system in the desired manner. As the controller moves the point $G(0.4j)$ of the nyquist curve to a point of $K(j\omega)G(j\omega)$ on the unit circle which has phase $130^0$ i.e. the desired phase margin of $50^0$ is achieved. In order to improve the closed loop performance the desired slope of the open loop Nyquist curve at the crossover frequency is set as $65^0$ thus reducing the present slope by $25^0$ which ensures a greater distance of the nyquist curve from the critical point at high frequencies. The controller parameters are found in two different approaches. At first the parameters are found using (15),(16),(17) where the slopes $s_a(\omega_c)$, $s_p(\omega_c)$ are approximated using (22),(23) which directly follows from bode integral. And, in this method the controller is found to be
$$K(s) = 1.3981(1 + \frac{1}{3.04s} + 1.37s) \quad (28)$$
The step response of the closed loop system thus found is shown in fig.(1)

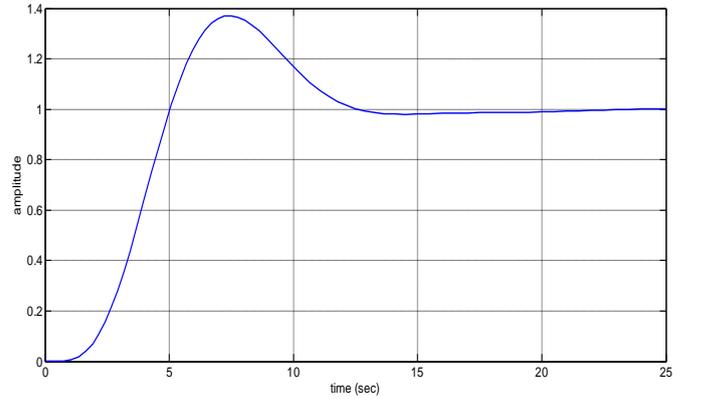

Figure 1. System response with delay directly using Bode integral

A second method based on Pade approximation is being proposed where the system with pure time delay can be approximated by a higher order transfer function free from delay term. This delay free transfer function is then used to find the step response of the closed loop system. On approximating the plant under consideration (27) using the first order pade approximation the approximated transfer function is found to be
$$G(s) = \frac{-s+20}{s^6 + 25s^5 + 110s^4 + 210s^3 + 205s^2 + 101s + 20} \quad (29)$$

On arriving at a rational transfer function free from time delay the controller parameters can easily be found using (15),(16),(17) directly to (29) and thus obtaining the parameters ensures improved performance in the desired manner . The slopes $s_a(\omega_c)$ , $s_p(\omega_c)$ are found from (10),(12) directly. However these approximations employed in order to obtain the slopes led to resultant Nyquist slope of $74^0$ i.e. almost a 13% error. However, the result demonstrates a significant improvement in the closed loop performance. The parameters of the PID controller required to manipulate the system specifications like crossover frequency, phase margin, Nyquist slope to the close loop response to predetermined desired values in order to obtain better closed loop performance. The PID controller thus found here is:

$$K(s) = 1.3726(1+\frac{1}{2.86s}+1.3327s) \qquad (30)$$

The step response of thus is shown in fig.(2)

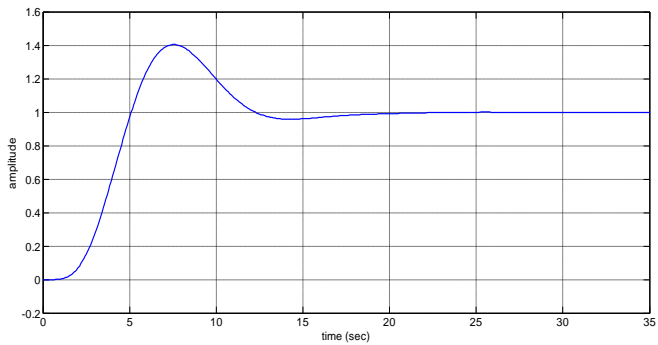

Figure 2. System response with delay by Pade aprroximation with PID controller via Bode integral

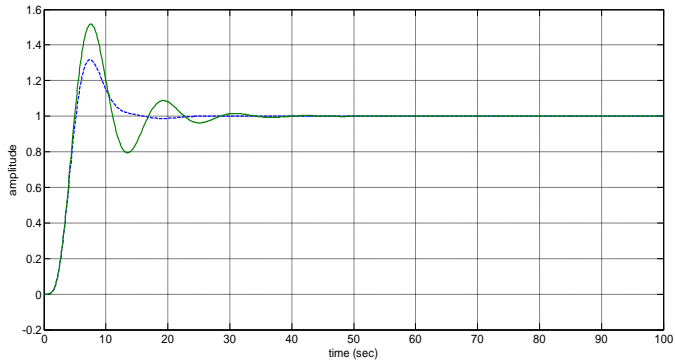

Figure 3. Comraison of closed loop System responses in Fig1 and Fig2

The system performance found using both the methods are compared with the response using a real coded Genetic algorithm(GA) in order to coming to a proper comparison for the best performance under the pre-chosen design specifications. A solution vector of the decision variables $\{K_p, K_i, K_d\}$ is initially randomly chosen from the search space and undergoes reproduction, crossover and mutation, in each iteration to give rise to a better population of solution vectors in further iterations [11] Instead of simple error minimization criteria for PID controller tuning the well known Integral of Time multiplied Absolute Error (ITAE) has been taken as the performance index .

$$J = \int_0^\infty t|e(t)|dt = \int_0^\infty t|r(t)-y(t)|dt$$

Tuning of the PID controller gains have been done in this study using the widely used population based optimizer known as Genetic Algorithm. Here, the number of population members in GA is chosen to be 50. The crossover and mutation fraction are chosen to be 0.9 and 0.3 respectively. The upper bound and lower bound of PID controller parameters are chosen in a way that the error in the resultant slope of the Nyquist curve does not exceed 20%. As for the parameters obtained by GA the error is almost 17% which is a 4% increase as compared to the method described in [6-8] due to the approximations of the slopes $s_a(\omega_c)$ , $s_p(\omega_c)$ . However as can be found from the comparative study that it leads to a much improved closed loop performance. The controller parameters found by the population base GA employed here are as follows,

$K_p = 1.3473, K_i = 0.3758, K_d = 1.875$

The PID controller thus found is:

$$K(s) = 1.3473(1+\frac{1}{3.58s}+1.3916s) \qquad (31)$$

And the corresponding step response is given in fig.(4)

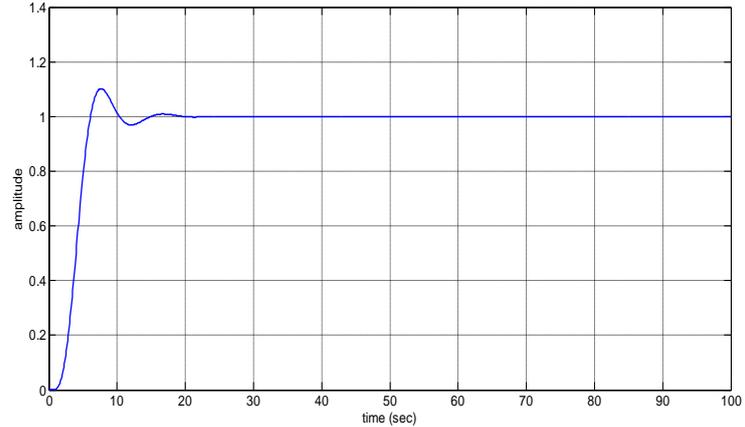

Figure 4. closed loop System response with delay based on GA

A proper comparison of the step responses of all the three mentioned PID tuning methodologies are shown in fig.(5)

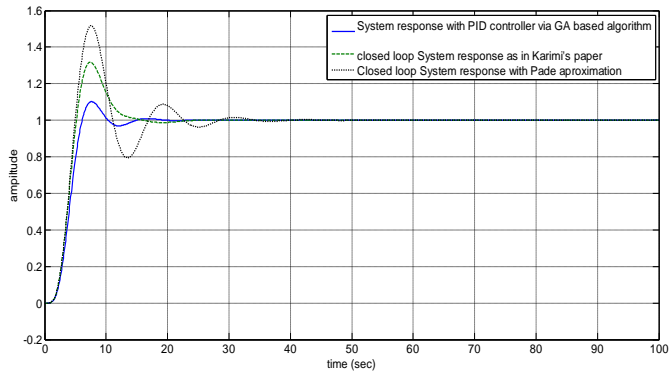

Figure5. Closed loopSystem response with delay by Bode integral as in Karimi's Paper , Pade approximation and Genetic algorithm based optimization

VII. CONCLUSION

From the comparison of the three methods it can be observed that the PID controller parameters tuned using the GA gives a much better performance than both of the other two methods proposed here. The PID controller parameters found from the pade approximated system shows much more overshoot as compared to the other two methods. Hence, it may be concluded that the GA based optimization results to a much improved closed loop performance of the system. However it is a trade off design because of the fact that improved closed loop performance led to an increased error in the slope of the resultant Nyquist curve. Further work includes development of an algorithm which improves the system performance with less amount of error in the resulting slope of the Nyquist curve.